\documentclass[sigconf]{acmart}
\AtBeginDocument{%
  }

\setcopyright{acmlicensed}
\copyrightyear{2018}
\acmYear{2018}
\acmDOI{XXXXXXX.XXXXXXX}

\acmConference[Conference acronym 'XX]{Make sure to enter the correct
  conference title from your rights confirmation emai}{June 03--05,
  2018}{Woodstock, NY}
\acmISBN{978-1-4503-XXXX-X/18/06}

\usepackage{multirow}
\usepackage{graphicx}
\usepackage[normalem]{ulem}
\useunder{\uline}{\ul}{}

\newcommand{\model}{GenRec}%

\begin{document}

\title{GenRec: Generative Sequential Recommendation with Large Language Models}

\author{Panfeng Cao}
\email{panfengc@umich.edu}
\affiliation{%
  \institution{University of Michigan}
  \city{Ann Arbor}
  \country{USA}}

\author{Pietro Liò}
\email{pl219@cam.ac.uk}



\begin{abstract}
Sequential recommendation is a task to capture hidden user preferences from historical user item interaction data and recommend next items for the user. Significant progress has been made in this domain by leveraging classification based learning methods. Inspired by the recent paradigm of ``pretrain, prompt and predict'' in NLP, we consider sequential recommendation as a sequence to sequence generation task and propose a novel model named Generative Recommendation (GenRec). Unlike classification based models that learn explicit user and item representations, GenRec utilizes the sequence modeling capability of Transformer and adopts the masked item prediction objective to effectively learn the hidden bidirectional sequential patterns. Different from existing generative sequential recommendation models, GenRec does not rely on manually designed hard prompts. The input to GenRec is textual user item sequence and the output is top ranked next items. Moreover, GenRec is lightweight and requires only a few hours to train effectively in low-resource settings, making it highly applicable to real-world scenarios and helping to democratize large language models in the sequential recommendation domain. Our extensive experiments have demonstrated that GenRec generalizes on various public real-world datasets and achieves state-of-the-art results. Our experiments also validate the effectiveness of the the proposed masked item prediction objective that improves the model performance by a large margin.
\end{abstract}



\received{10 May 2024}

\maketitle

\section{Introduction}
Recent years have witnessed the great success of recommender systems in online platforms such as Amazon and Yelp, which help people make micro decisions and fulfil their demands in daily life. Within online platforms, historical sequential user interaction data is utilized to capture the evolved and dynamic user behaviors and make appropriate recommendations of next items for the user. Different from traditional recommender systems that treat each user behavior as a sample and directly model the user preference on a single item, sequential recommendation learns timestamp-aware sequential patterns to recommend the next item \cite{10.1145/3568022}. For example, one might purchase peripheral accessories after buying a desktop. Various methods have been proposed to accurately model sequential user behaviors. \cite{hidasi2015session, 10.1145/3109859.3109877, 10.1145/3269206.3271761, 10.1145/3018661.3018689, 10.1145/2911451.2914683} employ Recurrent Neural Networks (RNN) to encode the left-to-right context. \cite{Caser} utilizes Convolutional Neural Networks (CNN) to learn sequential patterns as local features. \cite{10.1145/3404835.3462968} leverages Graph Neural Networks (GNN) to construct the item-item interest graphs and user interest sequences. \cite{S3-rec} designs contrastive loss to maximize the mutual information between different views of the sequential data. Contextual information such as item attributes is incorporated as self-supervised signals to learn user and item representations. 
The recent success of Transformer \cite{NIPS2017_3f5ee243} based large language models (LLMs) in various NLP tasks has inspired the research of utilizing LLMs in recommender systems \cite{SASRec, Bert4Rec, P5}. \cite{SASRec} utilizes the left-to-right self-attention mechanism to identify next relevant items from the sequential interaction history and \cite{Bert4Rec} further improves the performance by employing the bidirectional self-attention mechanism. Although both methods are effective, they do not consider personalization and the model performance is potentially limited. Generative LLMs based recommender system is considered a promising approach \cite{liu2023pre}. \cite{P5} is the first model to unify different recommendation tasks in the same framework to facilitate the knowledge transfer and it is pretrained across task-specific datasets to capture the deep semantics for personalization and recommendation. However, different modality information is encoded in the textual format of natural language, which might cause suboptimal modality representation. Moreover, \cite{P5} relies on manually designed task-specific hard prompts to formulate the problem as question answering, which requires additional data processing and prompt search.

\begin{figure*}
    \centering
    \includegraphics[scale=0.378]{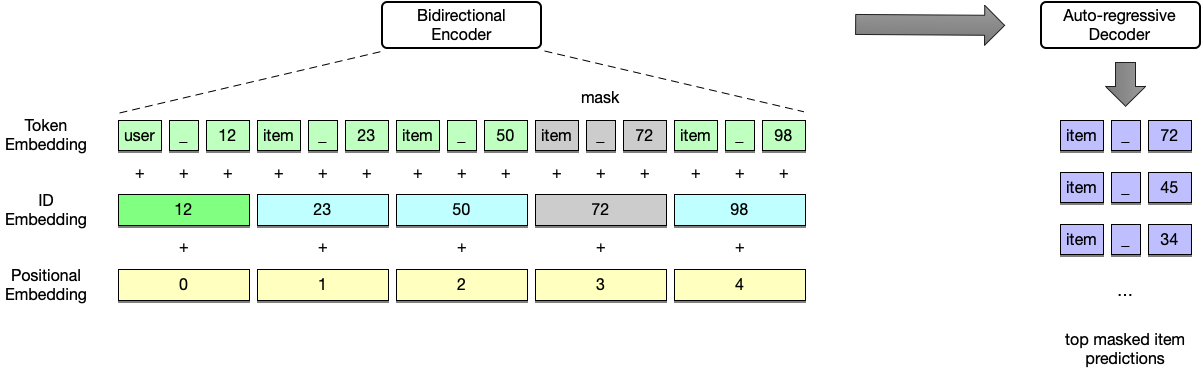}
    \caption{An illustration of the architecture of GenRec. The input textual user item interaction sequence is first tokenized into a sequence of tokens. Token embedding, ID embedding and positional embedding are summed up to produce the bidirectional encoder input. In pretraining and finetuning, a random item is masked and the auto-regressive decoder generates the masked item. In inference, the decoder generates top 20 masked item predictions to calculate the evaluation metrics.
    }
    \centering
    \label{fig:GenRec-arch}
\end{figure*}

To address the mentioned limitations, we propose GenRec, a novel generative framework for personalized sequential recommendation. GenRec utilizes the encoder-decoder Transformer as the backbone and formulates sequential recommendation as a sequence-to-sequence generation task. We utilize the cloze task \cite{devlin-etal-2019-bert} as the training objective and pretrain the model on the corpus to learn the bidirectional sequential patterns. Inspired by \cite{Bert4Rec}, to make the downstream sequential recommendation task consistent with the pretraining cloze task, we append the \texttt{[MASK]} token at the end of the input sequence to predict the next item. Extensive experiments on public real-world datasets demonstrate that GenRec generalizes effectively and achieves state-of-the-art performance across all datasets.
In conclusion, our contributions include: 
\begin{itemize}
\item We propose GenRec, a novel generative model for personalized sequential recommendation that can generate next items auto-regressively without prompts.
\item As far as our knowledge, it is the first generative sequential recommendation method that adopts the cloze objectives in both pretraining and finetuning. Our method only utilizes task specific datasets and does not rely on additional large pretraining corpus~\footnote{Our source code, datasets and pretrained models are publicly available at \url{https://github.com/caop-kie/GenRec}.}.
\item We conduct experiments on three public real-world datasets, demonstrating consistent improvements on multiple evaluation metrics compared with baseline methods.
\end{itemize}

\section{Related Work}
Transformer based LLMs have achieved remarkable success on various research fields such as text summarization, question answering, document understanding, etc. Typically, LLMs are trained on a vast amount of textual corpus from diverse sources including wikipedia, news, articles and books. LLMs can have emergent zero-shot learning capability as the model parameter size scales up with large training datasets \cite{zhao2023survey}, making LLMs better generalize to unseen domains. Recent efforts have been made to explore the potential of LLMs as recommender systems. \cite{P5} employs T5 as the model backbone and achieves competitive performance on a variety of recommendation tasks, demonstrating the generalization capability of LLMs based recommender systems. \cite{cui2022m6} develops a unified foundation model based on M6 and reduces the model size and training time by utilizing prompt tuning. Similar to \cite{P5}, user interaction data is represented as plain texts and recommendation tasks are formulated as either natural language understanding or generation. \cite{gao2023chat} leverages ChatGPT to build a conversational recommender system to improve the interactivity and explainability of the recommendation process. \cite{zhang2023recommendation} finetunes T5 with user-personalized instruction data, which is generated from manually designed templates, enabling users to communicate with the system with natural language instructions. \cite{10.1145/3604915.3608857} proposes an efficient framework to finetune LLMs for recommendation tasks and demonstrates significant improvements in the domains of movie and book recommendations. \cite{chen2023palr} employs a funneling approach where it first retrieves the candidates utilizing the user item interactions and then generates recommended items from candidates with a LLMs based framework.

\section{Task Formulation}
The goal of sequential recommendation is to predict the next item that the user is most likely to interact with given the historical item interaction sequence of the user. Formally, $i_{th}$ user $u_{i}$ has item interaction sequence $t_{i} = \{t^{i}_{1}, ..., t^{i}_{n_{i}}\}$ ordered chronologically. $t^{i}_{j}$ and $n_{i}$ denotes the $j_{th}$ item in the sequence and the length of $t_{i}$ respectively. $t^{i}_{j:k}$ denotes the subsequence, i.e., $t^{i}_{j:k} = \{t^{i}_{j}, t^{i}_{j+1}, ..., t^{i}_{k}\}$, where $1 \leq j < k \leq n_{i}$. Given the user item sequence $\{u_{i}, t^{i}_{1:n_{i}}\}$, the task of sequential recommendation is to predict the next item that $u_{i}$ is most likely to interact with at the $n_{i}+1$ timestep. 

\begin{figure*}
    \centering
    \includegraphics[scale=0.36]{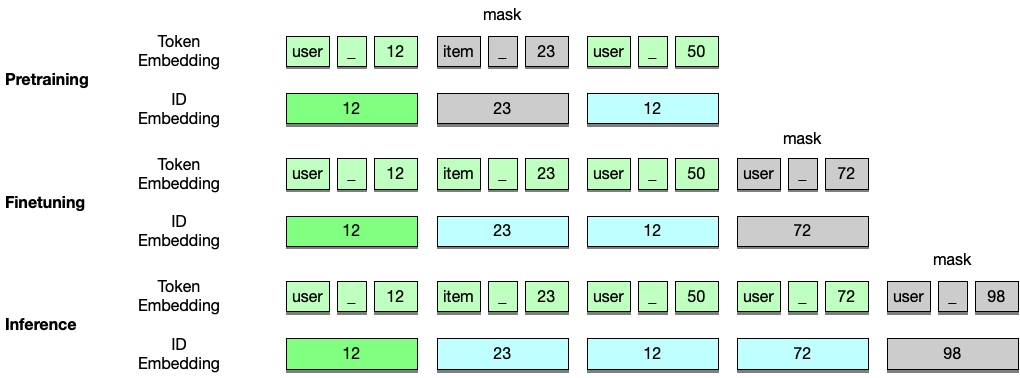}
    \caption{
    An illustration of different masking mechanisms in pretraining, finetuning and inference. In pretraining, a random item in the sequence is masked while in finetuning and inference, masked items are appended to the end of the sequence. Note, the last two items in the user item interaction sequence are excluded in pretraining to avoid data leakage. Similarly, the last one item in the sequence is excluded in finetuning.
    }
    \centering
    \label{fig:GenRec-mask}
\end{figure*}

\section{Methodology}
The overall architecture of \model{} is shown in Figure \ref{fig:GenRec-arch}. Our \model{} is established upon a sequence-to-sequence Transformer encoder-decoder framework. The \textit{encoder} of \model{} embeds cross modal user and item features from the input sequence and the \textit{decoder} generates next items auto-regressively. In the following sections, we elaborate the feature embedding and learning objectives of pretraining and finetuning.
\subsection{Model}
The feature embedding consists of token embedding, positional embedding, user ID embedding and item ID embedding. The input sequence to the model $\{u_{i}, t^{i}_{1:n_{i}}\}$ is tokenized into a sequence of tokens, which is wrapped around with the start indicator token [BEG] and the end indicator token [END]. Extra [PAD] tokens are appended to the end to unify the sequence length in the batch. The token sequence $S$ is represented as: 
\begin{equation}
    S = [BEG], Tok(u_{i}), Tok(t^{i}_{1}), ..., Tok(t^{i}_{n_{i}}), [END], ..., [PAD],
\end{equation}
where $Tok$ is the tokenizer function.
For the $j_{th}$ token in $S$, we apply textual token embedding, positional embedding and user or item ID embedding. The textual token embedding is formulated as $\textsc{Emb}(S_{j})$, where $\textsc{Emb}$ is the token embedding function. The positional information of tokens in the sequence is captured by the positional embedding $\textsc{PosEmb}(j)$, where $\textsc{PosEmb}$ is the 1D positional embedding function. And we utilize the user ID and item ID embedding to add personalized information of users and items to the sequence. Personalized information helps reveal the sequential patterns of user behaviors. For example, a user that previously viewed a Samsung tablet is more likely to continue viewing electronic products from Samsung or other brands. Different from P5 \cite{P5}, which apply whole-word embeddings shared by both user and item tokens. Our ID embedding layers are not shared to better capture modality specific features. The user ID embedding is represented by $\textsc{UIDEmb}(\textsc{UID}(j))$ and the item ID embedding is represented by $\textsc{TIDEmb}(\textsc{TID}(j))$. $\textsc{UIDEmb}$ and $\textsc{TIDEmb}$ are the ID embedding functions for users and items respectively. Since a word in the sequence can be split into multiple tokens after tokenization, we utilize $\textsc{UID}(j)$ or $\textsc{TID}(j)$ to retrieve the user ID or item ID for the $j_{th}$ token. For example, \textbf{item\_1234} is split into 4 tokens i.e. \textbf{item}, \textbf{\_}, \textbf{12} and \textbf{34}. Those tokens share the same item ID \textbf{1234} and embedding. Note only user tokens have user ID embeddings and only item tokens have item ID embeddings. All the aforementioned cross modal embeddings are added element-wisely to produce the final embedding $X$. Specifically the $j_{th}$ token cross modal embedding is formulated as: 
\begin{equation}
\begin{split}
    X_{j} = &\textsc{Emb}(S_{j}) + \textsc{PosEmb}(j) + \textsc{IDEmb}(\textsc{ID}(j)), j \in [0, n_{i}),\\
            &IDEmb \in \{UIDEmb, TIDEmb\},\\ 
            &ID \in \{UID, TID\}
\end{split}
\end{equation}

The decoder input is the next item that serves as the learning target. The same tokenizer and textual embedding in the encoder are utilized. Following baseline methods, the model generates top 20 items with beam search during inference to calculate the evaluation metrics.
\begin{table*}
\centering
\resizebox{\textwidth}{!}{%
\begin{tabular}{lllllllllllllllll}
\toprule
\multirow{2}{*}{Models} & \multicolumn{4}{c}{Sports}                                            & \multicolumn{4}{c}{Beauty}    & \multicolumn{4}{c}{Yelp}                                              \\ \cline{2-13} 
                        & HR@5            & NDCG@5          & HR@10           & NDCG@10         & HR@5            & NDCG@5          & HR@10           & NDCG@10         & HR@5            & NDCG@5          & HR@10           & NDCG@10 \\ \midrule
Caser                   & 0.0116          & 0.0072          & 0.0194          & 0.0097          & 0.0205          & 0.0131          & 0.0347          & 0.0176          & 0.0151          & 0.0096          & 0.0253          & 0.0129          \\ \hline
HGN                     & 0.0189          & 0.0120          & 0.0313          & 0.0159          & 0.0325          & 0.0206          & 0.0512          & 0.0266          & 0.0186          & 0.0115          & 0.0326          & 0.0159          \\ \hline
FDSA                    & 0.0182          & 0.0122          & 0.0288          & 0.0156          & 0.0267          & 0.0163          & 0.0407          & 0.0208          & 0.0158          & 0.0098          & 0.0276          & 0.0136          \\ \hline
GRU4Rec                 & 0.0129          & 0.0086          & 0.0204          & 0.0110          & 0.0164          & 0.0099          & 0.0283          & 0.0137          & 0.0152          & 0.0099          & 0.0263          & 0.0134          \\ \hline
BERT4Rec                & 0.0115          & 0.0075          & 0.0191          & 0.0099          & 0.0203          & 0.0124          & 0.0347          & 0.0170          & 0.0051          & 0.0033          & 0.0090          & 0.0045          \\ \hline
P5-S                    & 0.0272          & 0.0169          & 0.0361          & 0.0198          & 0.0503          & 0.0370          & {\ul 0.0659}    & 0.0421          & {\ul 0.0568}    & {\ul 0.0402}    & {\ul 0.0707}    & {\ul 0.0447}    \\ \hline
P5-B                    & {\ul 0.0387}    & {\ul 0.0312}    & {\ul 0.0460}    & {\ul 0.0336}    & {\ul 0.0508}    & {\ul 0.0379}    & \textbf{0.0664} & {\ul 0.0429}    & -               & -               & -               & -               \\ \hline
SASRec                  & 0.0233          & 0.0154          & 0.0350          & 0.0192          & 0.0387          & 0.0249          & 0.0605          & 0.0318          & 0.0162          & 0.0100          & 0.0274          & 0.0136          \\ \hline
S3-Rec                  & 0.0251          & 0.0161          & 0.0385          & 0.0204          & 0.0387          & 0.0244          & 0.0647          & 0.0327          & 0.0201          & 0.0123          & 0.0341          & 0.0168          \\ \hline
GenRec                  & \textbf{0.0397} & \textbf{0.0332} & \textbf{0.0462} & \textbf{0.0353} & \textbf{0.0515} & \textbf{0.0397} & 0.0641          & \textbf{0.0439} & \textbf{0.0627} & \textbf{0.0475} & \textbf{0.0724} & \textbf{0.0507} 
\\ \bottomrule
\end{tabular}%
}
\caption{Performance comparison between GenRec and baselines on Sports, Beauty and Yelp datasets.}
\label{tab:metrics}
\end{table*}

\subsection{Pretraining}
\label{sec:pretraining}
We pretrain \model{} with the masked sequence modeling task to deeply fuse the personalized sequential patterns. Given a user item sequence, we randomly sample an item, which is not necessarily the last item, from the sequence and replace it with the \texttt{[MASK]} token. The model is trained to generate the masked item in the decoder output. Cross-entropy loss is computed between the generated item and the masked item and minimized in pretraining. The masking process is also illustrated in Figure \ref{fig:GenRec-mask}. Different from \cite{Bert4Rec}, where a number of items are masked, we only mask a single item in the sequence to unify the learning objectives of pretraining and finetuning. Our bidirectional masked sequence modeling task is based on the observation that the item sequence is not strictly ordered. For example, given similar items item\_1 and item\_2, user\_1 might interact with item\_1 and then item\_2 while user\_2 interacts with item\_2 and then item\_1. It is crucial to incorporate both left and right contexts to encode the user behavior \cite{Bert4Rec}. To avoid the data leakage, we pretrain the model on the training split of datasets (See \S \ref{sec:datasets} for details about the training split).
\subsection{Finetuning}
In the finetuning stage, the learnt user and item sequential patterns in pretraining are utilized for the next item prediction task. As mentioned in section \ref{sec:pretraining}, finetuning has the same learning objective with pretraining for knowledge transfer. Following the masking mechanisms in Figure \ref{fig:GenRec-mask}, we prepare the dataset in the same format as in pretraining by appending the \texttt{[MASK]} token to the end of the token sequence. The input to the model is the masked user item sequence and the output is the predicted next item.
\section{Experiments}

\section{Datasets}
\label{sec:datasets}
Amazon Sports and Amazon Beauty datasets are obtained from Amazon review datasets \cite{amazon_datasets}, which are collected from Amazon.com online platform. 
Yelp dataset is a large business recommendation dataset. Following \cite{S3-rec, P5}, we use transactions between January 1, 2019 and December 31, 2019. For all datasets, item sequences are extracted per user and sorted ascendingly by the interaction timestamps. Following the same data preprocessing in the baseline methods, unpopular items and inactive users with less than five interaction records are filtered. The statistics of the datasets after preprocessing are summarized in Table \ref{tab:datasets}.

To generate the training, validation and test splits of datasets, we apply the leave-one-out strategy following \cite{P5, S3-rec, SASRec, Bert4Rec}. In each user item sequence, we use the last item as the test data, the item before the last item as the validation data, and the remaining item sequence as the training data.

\begin{table}
\centering
\resizebox{0.35\textwidth}{!}{%
\begin{tabular}{llll}
\toprule
\textbf{Dataset} & \textbf{Sports} & \textbf{Beauty} & \textbf{Yelp} \\ \midrule
\# Users                  & 35,598      & 22,363 & 30,431 \\ \hline
\# Items                  & 18,357      & 12,101 & 20,033 \\ \hline
Avg. Items / User         & 8.3         & 8.9    & 10.4   \\ \hline
Avg. Users / Item         & 16.1        & 16.4   & 15.8   \\ \bottomrule
\end{tabular}%
}
\caption{Statistics of the datasets.}
\label{tab:datasets}
\end{table}

\section{Baselines}
\label{sec:baselines}
We compare the performance of GenRec with several competitive sequential recommendation baselines.

Caser \cite{Caser} is a CNN based method, which utilizes both horizontal and vertical convolutional filters to capture high-order Markov chains for sequential recommendation.

FDSA \cite{FDSA} models the feature sequence with the self attention module and captures the feature transition patterns.

GRU4Rec \cite{hidasi2016sessionbased} applies GRU to model the user click history sequence for session based recommendation.

HGN \cite{HGN} adopts hierarchical gating networks to learn both long-term and short-term user behaviors.

Bert4Rec \cite{Bert4Rec} learns the bidirectional item representation by masked language modeling for sequential recommendation.

SASRec \cite{SASRec} applies the self attention mechanism to model the user sequential behaviors.

S3Rec \cite{S3-rec} designs self-supervised learning objectives to capture the correlations of items and attributes for sequential recommendation.

P5 \cite{P5} is a pretrained language model for recommendation and unifies different recommendation tasks by formulating them in natural language with prompts.

All the baselines except P5 are classification based.

\subsection{Implementation Details}
The model weights of \model{} is initialized from the pretrained BART \cite{lewis-etal-2020-bart} base model, which consists of a 6-layer Transformer encoder and a 6-layer Transformer decoder. There are 12 attention heads for both encoder and decoder and the model dimension is 768. The total number of model parameters is 184 million. We use the pretrained BART tokenizer for tokenization. \model{} is trained with an AdamW optimizer \cite{adamW}, in which 1e-5 is set as the learning rate and weight decay. The maximum length of the input tokens is set to 512. First 5\% of iterations are used as the warmup stage. The beam size is set to 20 in inference. \model{} is pretrained for about an hour and finetuned for 25 epochs on one NVIDIA RTX 3090 GPU.

\subsection{Evaluation Metrics}
To evaluate the model performance, we employ top-k Hit Ratio (HR@k) and Normalized Discounted Cumulative Gain (NDCG@k) and report the the results of HR@\{1, 5, 10\} and NDCG@\{5, 10\} under the all-item setting, i.e. all items are possible candidates to be recommended as the next item. Our model is evaluated on Amazon Sports, Amazon Beauty and Yelp datasets. In the tables, \textbf{bold} numbers refer to the best scores, while \underline{underlined} numbers refer to the second best scores.

\subsection{Results and Analysis}
This section provides experimental results and analyses of our model's effectiveness. The performance metrics of the baselines are obtained from the public results in \cite{P5}. See Appendix~\ref{sec:baselines} for more details about baselines. As is presented in Table \ref{tab:metrics}, GenRec outperforms other baselines on all metrics on Sports and Yelp datasets. Although it is outperformed by P5 on the HR@10 metrics of the Beauty dataset, GenRec still achieves comparable performance with other baselines. 

We speculate due to the smaller size of Beauty dataset, the sequential patterns are not learnt thoroughly in pretraining. Increasing the number of pretraining epochs could potentially improve the performance. Since we focus on low-resource efficient training in this work, we leave the improvement for future work. Since P5 is a unified model and trained using different types of recommendation tasks, it learns better user item representation with the knowledge transfer between tasks on smaller datasets. The advantage of our method compared with P5 is that our model is lightweight and the training only takes a few hours to complete. We do not leverage prompt engineering and no prompt search is required to find the best prompt. 

\begin{table}[]
\centering
\resizebox{0.49\textwidth}{!}{%
\begin{tabular}{lllll}
\toprule
\multicolumn{1}{c}{\textbf{Dataset}}          & \multicolumn{1}{c}{\textbf{HR@5}}    & \multicolumn{1}{c}{\textbf{NDCG@5}} & \multicolumn{1}{c}{\textbf{HR@10}} & \multicolumn{1}{c}{\textbf{NDCG@10}}                    \\ \midrule
Sports     & \textbf{0.0397}  & \textbf{0.0332} & \textbf{0.0462} & \textbf{0.0353} \\ 
Sports w/o pretraining     & 0.0360  & 0.0286 & 0.0431 & 0.0310 \\ \hline
Beauty            & \textbf{0.0515} & \textbf{0.0397} & \textbf{0.0641} & \textbf{0.0439} \\ 
Beauty w/o pretraining     & 0.0422 & 0.0313 & 0.0548 & 0.0354 \\ \hline
Yelp             & \textbf{0.0627} & \textbf{0.0475} & \textbf{0.0724} & \textbf{0.0507} \\ 
Yelp w/o pretraining       & 0.0626 & 0.0469 & 0.0716 & 0.0499 \\ \bottomrule
\end{tabular}
}
\caption{Ablation study of our method on four datasets. Pretraining effectively improves the model performance across all datasets.}
\label{tab:ab-study}
\end{table}

\section{Ablation Study}
To verify the effectiveness of the proposed masked sequence modeling task, we conduct the ablation study on Amazon Sports, Amazon Beauty and Yelp datasets. As is shown in Table \ref{tab:ab-study}, the model performance drops across all metrics without the masked sequence modeling task. It indicates the pretraining objective helps the model effectively capture the user behavior patterns and learn the user item representations. It is worth mentioning even without pretraining, GenRec still achieves comparable performance with P5 and outperforms other baselines, which proves the effectiveness of our method.

\section{Conclusion}
In this work, we propose GenRec, a novel sequence to sequence framework that generates personalized sequential recommendation. Compared with existing generative models, GenRec is lightweight and efficient and does not rely on prompt engineering to find the best prompt. By leveraging the Transformer architecture as the backbone, GenRec effectively learns the bidirectional sequential patterns with the attention mechanism and achieves state-of-the-art performance on various public datasets. Besides, our proposed method is also flexible to integrate with other sequence to sequence language models and can be applied in other recommendation domains such as direct recommendation, which we leave for future work.

\bibliographystyle{ACM-Reference-Format}
\bibliography{GenRec}
\end{document}